\documentclass{ws-p8-50x6-00}
\usepackage{amssymb,amsmath,amsfonts,latexsym}
\input epsf
\begin{document}
\title{Search for Critical Phenomena in Pb+Pb Collisions}
\author{Mikhail L. KOPYTINE\thanks{\protect{
on an unpaid leave from P.N.Lebedev Physical Institute,
Russian Academy of Sciences}}\ \ for The NA44 Collaboration}
\address{
Department of Physics and Astronomy, SUNY at Stony Brook, 
\\E-mail: Mikhail.Kopytine@sunysb.edu}
\author
{I.Bearden$^{a}$, H.B{\O}ggild$^{a}$, J.Boissevain$^{b}$, L.Conin$^{d}$,
J.Dodd$^{c}$, B.Erazmus$^{d}$, S.Esumi$^{e}$, C.W.Fabjan$^{f}$, 
D.Ferenc$^{g}$, D.E.Fields$^{b}$, A.Franz$^{f}$, J.J.Gaardh{\O}je$^{a}$,
A.G.Hansen$^{a}$, O.Hansen$^{a}$, D.Hardtke$^{i}$, H. van Hecke$^{b}$, 
E.B.Holzer$^{f}$, T.J.Humanic$^i$, P.Hummel$^f$, B.V.Jacak$^j$, R.Jayanti$^i$,
K.Kaimi$^e$, M.Kaneta$^e$, T.Kohama$^e$, M.L.Kopytine$^j$, M.Leltchouk$^c$,
A.Ljubicic, Jr$^g$, B. L{\"o}rstad$^k$, N.Maeda$^e$, L.Martin$^d$, 
A.Medvedev$^c$, M.Murray$^h$, H.Ohnishi$^e$, G.Paic$^f$, S.U.Pandey$^i$,
F.Piuz$^f$, J.Pluta$^d$, V.Polychronakos$^l$, M.Potekhin$^c$, G.Poulard$^f$,
D.Reichhold$^i$, A.Sakaguchi$^e$, J.Schmidt-S{\O}rensen$^k$, J.Simon-Gillo$^b$,
W.Sondheim$^b$, T.Sugitate$^e$, J.P.Sullivan$^b$, Y.Sumi$^e$, W.J.Willis$^c$,
K.L.Wolf$^h$, N.Xu$^b$, D.S.Zachary$^i$.}
\address{$^A$ Niels Bohr Institute, Denmark;
$^B$ LANL, USA;
$^C$ Columbia U., USA;
$^D$ Nuclear Physics Laboratory of Nantes, France;
$^E$ Hiroshima U., Japan;
$^F$ CERN;
$^G$ Rudjer Boscovic Institute, Croatia;
$^H$ Texas A\&M U., USA;
$^I$ The Ohio State U., USA;
$^J$ SUNY at Stony Brook, USA;
$^K$ U. of Lund, Sweden;
$^L$ BNL, USA.
}

\maketitle
\abstracts{ NA44 uses a 512 channel Si pad array covering 
$1.5 <\eta < 3.3$
to study charged hadron production in Pb+Pb collisions at
the CERN SPS.  We apply a multiresolution analysis, based on a Discrete Wavelet
Transformation, to probe the texture of particle distributions 
event-by-event,
by simultaneous localization of features in space and scale.  Scanning a
broad range of multiplicities, we look for a possible critical
behaviour in the power spectra of local density fluctuations.  The data
are compared with detailed simulations of detector response, using
heavy ion event generators, and with a reference sample 
created via event mixing.}

An ultrarelativistic collision of heavy ions presents a phenomenon whose
most interesting features are conditioned by the large multitude of
degrees of freedom involved, and yet offer an opportunity for the
fundamental physics of the strong interaction to manifest itself.
The very notion of a phase transition in such collisions is inherently
of a multiparticle nature.
Truly multiparticle observables, defined on event-by-event basis --
a few have been constructed so far -- therefore attract attention.
Recently published event-by-event analyses of the 158 GeV/A $Pb+Pb$ data
either deal with a small number of events \cite{EMU15}
or analyse properties of a large ensemble of events
using a single observable ($p_T$) 
to compare different ensemble averages\cite{NA49_phi}.
In the first case, 
accumulation of feature information from large data sets
remains open.
In the second case \cite{NA49_phi}, 
one can not establish a scale independency in event textures by observing
a logical consequence thereof
\cite{CLT}.
Furthermore, an ensemble average
on a set of \emph{post-freeze-out} events is not representative of the  
\emph{pre-freeze-out} history of those events, 
due to the dramatic non-stationarity of the open system, with a consequent
lack of ergodicity.

Here we concentrate on \emph{texture}, or \emph{local
fluctuation} observables, where 
a single event determines its own correlation/fluctuation content,
and the scale composition thereof manifests itself
in the observables in a positive way.
The idea to look at particle distributions in rapidity $y$ to search for
critical behaviour was proposed 
\cite{Scalapino_Sugar,Carruthers_Sarcevic}
based upon a
Ginzburg-Landau type of multihadron production theory
\cite{Scalapino_Sugar},
where a random hadronic field $\phi(y)$ plays the role of an order parameter
in a hadronization transition.
Enhanced large scale correlations of hadrons in $y$ at 
the phase transition
would signal critical fluctuations in the order parameter.
Stephanov and coworkers \cite{tricritical} indicated a second order
QCD phase transition point which should exist
under certain initial conditions, within the reach of today's
experiments.

In our work, 
a power spectrum analysis of event texture in pseudorapidity
$\eta$ and azimuthal angle $\zeta$,
based on a 
Discrete Wavelet Transformation (DWT)\cite{DWT}, is 
performed on a number of large event ensembles sampled according
to their multiplcity, thereby studying the impact parameter dependence
of the observables.
DWT quantifies contributions of different  $\zeta$ and $\eta$ scales
into the overall event's texture, thus testing the possible 
large scale enhancement.

The SPS beam was collimated to  a $1\times2$ mm profile.
The NA44 Si pad array, 
installed 10 cm downstream from the target, in the magnetic field of the
first dipole\cite{NA44ex},
measured ionization energy loss of charged particles
 in its 512 300 $\mu$m thick Si pads.
The silicon detector had 
inner radius 7.2 mm and outer radius 43 mm.
The detector
was split \emph{radially} into 16 
rings of equal 
$\eta$ coverage.
Each  ring was divided
\emph{azimuthally} into 32 sectors of equal angular coverage to form pads.
The pads were read out by AMPLEX 
\cite{AMPLEX}
chips, one chip per sector.
$\delta$-electrons, produced by the $Pb$ beam traversing
the target, were swept away to one side by the dipole magnetic field
($\le 1.6$ Tl).
Only the $\delta$-electron-free side 
was used in this analysis.
Channel pedestals had, on the average,  $FWHM=0.48 <dE>$
of 1 MIP.
In the texture analysis, every event was represented by a 2D array of
the calibrated digitized amplitudes of the channels ( an
\emph{amplitude array}).
Empty target runs were used to measure the background.
Cross-talk in the detector was evaluated off-line.

DWT formalizes the images
of the $PbPb$ collision events in pseudorapidity $\eta$ and azimuthal
angle $\zeta$ by expanding them into a set of functions orthogonal with
respect to scale and position, and allows one 
to accumulate the texture information 
by averaging the power spectra of many events.
The simplest DWT basis is the Haar one, built upon the scaling function
$\phi(x) = 1$ for $0\le x<1$ and 0 otherwise.
If the interaction vertex lies on the detector's symmetry axis,
every pad's acceptance is a rectangle in the $(\zeta,\eta)$ space.
Then, the Haar basis is the natural one, as its scaling function in 
two dimenstions (2D)
$\Phi(\zeta,\eta) = \phi(\zeta)\phi(\eta)$ 
is just a pad's acceptance (modulo units). 
We set up a 2D 
wavelet basis:
\begin{equation}
\Psi^{\lambda}_{m,i,j}(\zeta,\eta) =
 2^{m}\Psi^{\lambda}(2^{m}\zeta-i,2^{m}\eta-j)
 \label{dilate_translate}
\end{equation}
$\Phi_{m,i,j}(\zeta,\eta)$ is constructed from $\Phi(\zeta,\eta)$
similarly.
Here, $m$ is an integer scale fineness 
index; $i$ and $j$ index the discrete positions of pad centers
in $\zeta$ and $\eta$ 
(
$1 \le m \le 4$ and $1\le i,j \le 16$
because we use $16=2^4$ rings and 16 sectors
).
Different values of $\lambda$ 
(denoted as $\zeta$, $\eta$, and $\zeta\eta$) distinguish, respectively,
functions with  azimuthal, pseudorapidity, and diagonal texture 
sensitivity:
\begin{equation}
\Psi^\zeta=\psi(\zeta)\phi(\eta), \ \ 
\Psi^\eta=\phi(\zeta)\psi(\eta), \ \ 
\Psi^{\zeta\eta}=\psi(\zeta)\psi(\eta)
\end{equation}
In the Haar basis, for any variable $x$
\begin{equation}
\psi(x) =     \{ +1 \mbox{\ for\ } 0\le x<\frac{1}{2}; 
                 -1 \mbox{\ for\ }  \frac{1}{2}\le x<1;
                  0 \mbox{\ otherwise}
              \}    
\end{equation}
is the wavelet function.
Then, $\Psi^\lambda_{m,i,j}$ with integer $m$, $i$, and $j$ are known 
\cite{DWT}
to form an othonormal basis in $L^2({\mathbb{R}}^2)$.
We adopt the existing \cite{DWT_power} 1D DWT power spectrum analysis
technique  
and expand it to 2D.
Track density in an individual event is
$\rho(\zeta,\eta)$ and its \emph{local} fluctuation
in a given event is
$ \sigma^2 \equiv \langle \rho - \bar{\rho},\rho - \bar{\rho}\rangle,$
where $\bar{\rho}$ is the average $\rho$ in the given event.
Using completeness of the basis, we expand 
\begin{equation}
\rho - \bar{\rho} = 
\langle \rho,\Psi^\lambda_{m,i,j}\rangle \Psi^\lambda_{m,i,j}
- \langle \bar{\rho},\Psi^\lambda_{m,i,j}\rangle \Psi^\lambda_{m,i,j}
\end{equation}

Notice that  $\bar{\rho}$, 
being constant
within detector's 
rectangular acceptance, 
is orthogonal
to any  $\Psi^\lambda_{m,i,j}$ with $m \ge 1$.
Due to the
orthonormality condition 
$\langle \Psi^\lambda_{m,i,j},\Psi^{\lambda'}_{m',i',j'}\rangle = 
\delta_{\lambda,\lambda'}\delta_{m,m'}\delta_{i,i'}\delta_{j,j'}$,
the $\rho - \bar{\rho}$ components for
different scales do not form cross-terms in the $\sigma^2$ sum, 
and the sum 
contains no cross-terms between $\rho$ and $\bar{\rho}$ for the four
observable scales.
Instead of a $\langle \rho, \Phi_{m=5,i,j} \rangle$ set, the amplitude array 
-- its closest experimentally achievable approximation -- is used as the
DWT input. 

The Fourier images of 1D 
wavelet functions occupy a set of wave numbers 
whose characteristic broadness grows with scale fineness $m$ as $2^m$;
$2^{2m}$ should be used in the 2D case.
In 2D, we find it most informative to present the three modes of a power
spectrum 
with different directions of sensitivity
$P^{\zeta\eta}(m)$, $P^\zeta(m)$, $P^\eta(m)$
separately.
We define \emph{power spectrum} as 
\begin{equation}
P^\lambda(m) = 
\frac{1}{2^{2m}}\sum_{i,j}\langle \rho,\Psi^\lambda_{m,i,j}\rangle^2 ,
\label{eq:P_m}
\end{equation}
where the denominator gives the meaning of spectral \emph{density}
to the observable.
So defined, the $P(m)$ of a random white noise field (for example)
is independent of $m$ \cite{N_Wiener}.
In the first approximation, the white noise example provides a 
base-line case for comparisons in search
for non-trivial effects.

We used WAILI \cite{WAILI} software library to obtain the wavelet expansions.
\begin{figure}
\epsfysize=3.5cm
\centerline{\epsfbox{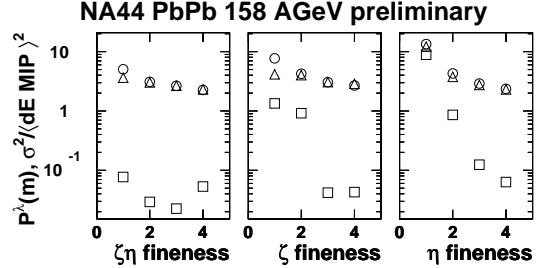}}
\caption{
Power spectra of $7\times10^3$ events in the multiplcity bin
$379<\,dN/\,d\eta<463$. 
$\bigcirc$ -- true events,
$\bigtriangleup$ -- mixed events,
$\protect\Box$ -- the average event.
}
\label{compare}
\end{figure}
Figure \ref{compare} 
shows the power spectra for one  multiplicity range.
The first striking feature is that the power spectra of physical events
are indeed enhanced on the coarse scale. 
The task of the analysis is to quantify and, as much as possible, eliminate
``trivial'' and experiment-specific reasons for this enhancement.
\begin{figure}[!ht]
\epsfysize=8.4cm
\centerline{\epsfbox{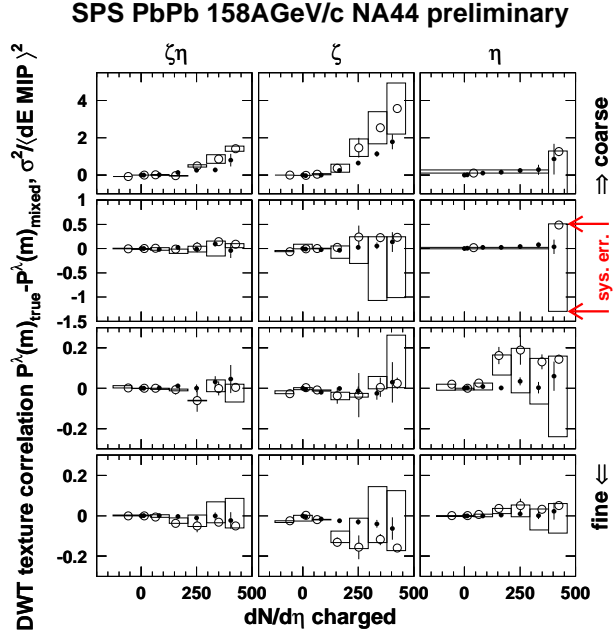}}
\caption{Multiplicity dependence of the texture correlation.
$\bigcirc$ -- the NA44 data, $\bullet$ -- RQMD. 
The boxes show the systematic errors vertically and the boundaries of
the multiplcity bins horizontally; the statistical errors 
are indicated by the vertical bars on the points. The rows correspond
to the scale fineness $m$, the columns -- to the directional mode
of the power spectrum $\lambda$ (which can be diagonal $\zeta\eta$,
azimuthal $\zeta$, and pseudorapidity $\eta$).
}
\label{multi_dep}
\end{figure}

The average event, formed by summing amplitude arrays of the
measured  events within a  multiplicity range,
and dividing by the number of events, has a much reduced texture as
fluctuations cancel.  
However it retains the
texture associated with the $\,d^2N/\,d\eta\,d\zeta$, with the dead
channels and the finiteness of the beam's geometrical cross-section.
A better way to get rid of the ``trivial'' texture is to use mixed
events.  
The event mixing is done by 
taking different channels from different events.  
Therefore, the
mixed events preserve the texture associated with 
the detector position offset,
the inherent $\,dN/\,d\eta$ shape and the dead channels.
In order to reproduce the electronics cross-talk effects
in the mixed event sample, mixing is done sector-wise, i.e. the
sectors constitute 
the subevents subjected to the event number scrambling.
In other words, the mixed events preserve the texture coupled with
the channels of the detector.
This is \emph{static} texture as it reproduces
its pattern event after event; 
we are interested in \emph{dynamic} texture.
We reduce sources of  static texture in the 
power spectra by empty target subtraction and by subtraction of the mixed
events power spectra, thus obtaining the \emph{texture correlation}
$P^\lambda(m)_{true} - P^\lambda(m)_{mixed}$.
Its multiplicity dependence is plotted on Figure \ref{multi_dep}.
For comparison with models, a MC simulation (done with RQMD \cite{RQMD}) 
includes the known static texture effects and undergoes the same elimination 
procedure. 
This allows the effects irreducible by the subtraction methods to be
taken into account in the comparison.
One such example is the finite beam size,
which has been shown by the MC studies to cause the RQMD points
to rise with $\,dN/\,d\eta$.

The systematic errors were evaluated by removing the $Pb$ target and 
switching magnetic field polarity to expose the analyzed side of the detector 
to $\delta$-electrons, while minimizing nuclear interactions.
All correlations (i.e. deviations of 
$P^\lambda(m)_{true}$ from $P^\lambda(m)_{mixed}$) 
in such events are considered to be systematic errors. 
Thus this component of the systematic error gets a sign, 
and the systematic errors are asymmetric. 
The other component (significant only on the coarsest scale) 
is the uncertainty of our knowledge of the beam's geometrical cross-section. 

This novel method of event-by-event analysis,
 applied to the SPS $PbPb$ data, does not reveal any evidence
of critical phenomena.

The authors thank N.Antoniou, I.Dremin, E.Shuryak, M.Stephanov, 
and T.Trainor for illuminating discussions.
The NA44 Collaboration thanks the staff of the CERN PS-SPS 
accelerator complex for their excellent work, and the technical staff in
the collaborating institutes for their valuable contributions.
This work was supported by the Austrian Fonds zur F{\"o}rderung der
Wissenschaftlichen Forschung;
the Science Research Council of Denmark;
the Japanese Society for the Promotion of Science; the Ministry of
Education, Science and Culture, Japan;  the Science Research Council 
of Sweden; the US Department of Energy and the National Science Foundation.

\end{document}